\documentclass[twocolumn,prl,showpacs,preprintnumbers,amsmath,amssymb]{revtex4}

\usepackage{graphicx}
\usepackage{dcolumn}
\usepackage{bm}

\begin{document}

\title{Dilute Hard ``Sphere'' Bose Gas in Dimensions 2, 4 and 5 }
\author{C. N. Yang}

\affiliation{Chinese University of Hong Kong, Hong Kong and
Tsinghua University, Beijing, China}

\begin{abstract}

The ground state energy for a dilute hard ``sphere" Bose gas in
various dimensions is studied theoretically.

\end{abstract}

\date{\today}
\maketitle




Consider a collection of $N$ Bose hard spheres in a periodic
$L\times L \times L$ box.  The ground state energy $E_0$ of the
system was given \cite{LHY} in an asymptotic expansion in 1957:
\begin{equation}
\frac{E_0}{N}= 4\pi a\rho [1+\frac{128}{15\sqrt{\pi}}\sqrt{\rho
a^3}+\ldots]
\end{equation}
in the limit that $N\rightarrow \infty$ at fixed density
$\rho=N/\Omega$, for small value of $\rho a^3$.  We follow the
notation of reference 1, now usually called LHY. At that time
equation (1) was a purely theoretical result.  Today with the
amazing developments of laser technology and computer power it has
become possible \cite{Exp}  to test the validity of the expansion
(1). In the present paper we attempt to find similar expansions
for the same problem in dimensions 2, 4 and 5.

For dimension 1, the problem had been solved in reference 3.

\section{Review of Dimension 3}

The one dimensional problem is special since there is no
diffraction,  only reflection,  in one dimension. In higher
dimensions diffraction is present.

For the ground state only S-wave scattering need be \cite{Huang}
considered in the limit of fixed $\rho$, $N\rightarrow \infty$ and
$a\rightarrow 0$. In this section, we review the derivation of
equation (1) in dimension 3.

The boundary condition that the wave function vanishes when
$|\textbf{r}_i-\textbf{r}_j|=0$ is mathematically simple to
define, but difficult to analyze. We shall therefore replace it by
a potential energy called the ``\emph{pseudopotential}" originally
\cite{Fermi} due to Fermi:
\begin{eqnarray}
&&H=-\sum_i \nabla_i^2+V \nonumber \\
&V&=4\pi a \int d^3 \textbf{r}_1 d^3
\textbf{r}_2\psi^*(\textbf{r}_1)\psi^*(\textbf{r}_2)\delta^3(\textbf{r}_1-\textbf{r}_2)\frac{\partial}{\partial
r_{12}}\nonumber \ \ \ [2]
\\&&\hspace{4cm}\times[\textbf{r}_{12}\psi(\textbf{r}_1)\psi(\textbf{r}_2)].\nonumber
\end{eqnarray}
All equations with a square bracket refer to equations in LHY.

To generalize to other dimensions we need to understand why the
pseudopotential can replace the boundary condition when two
spheres touch, at least in low orders of $a$.   Now it is known
from electrostatics that
\begin{equation}
\nabla^2 \frac{1}{r}=-4\pi \delta^3(\textbf{r}).
\end{equation}
Thus for 2 bodies
\begin{equation}
[-\nabla^2+4\pi a\delta^3(\textbf{r})\frac{\partial}{\partial
r}r]\varphi=4\pi a\delta^3(\textbf{r})[\frac{\partial}{\partial
r}r-1],
\end{equation}
if \begin{equation} \varphi=1-\frac{a}{r}.
\end{equation}
Now (a) the RHS of (3) when operating on any function that is not
singular at $r=0$ gives zero. (b) $\varphi$ satisfies the boundary
condition that $\varphi=0$ at $r=a$, so that $\varphi$ is the
correct $S$ wave scattered wave function for the collision of
particles 1 and 2. This means that the $N = 2$ problem can be
replaced by a pseudopotential
\begin{equation}
V=8\pi
a\delta^3(\textbf{r}_{12})\frac{\partial}{\partial{r_{12}}}r_{12},\nonumber
\hspace{4cm}\text{(PP3)}
\end{equation}
where we have taken into account the reduced mass for the 2 body
system.  For the $N$ body problem we thus obtain the potential $V$
of equation [2] in LHY.

The next step is to use the simpler potential $V'$ [3] in LHY
which operating on any nonsingular wave function gives the same
result as $V$.

With $V'$ LHY obtained for the ground state the first order
perturbation energy and wave function.  With $V'$ the second order
perturbation energy, however, diverges.  LHY showed that that is
due to the fact $V'$ is not $V$ in the second order calculation.
Using $V$ and not $V'$ in the second order calculation, LHY showed
that
\begin{equation}
E_0=4\pi a\rho N-\sum {}'[k^2+k_0^2-k
\sqrt{k^2+2k_0^2}-\frac{k_0^4}{2k^2}],\nonumber \hspace{1cm}[23]
\end{equation}
which is convergent and gave equation (1) above.

\section{DIMENSION 4}
To generalize to dimension 4 we need generalizations of equations
(2)$\rightarrow$(4):
\begin{equation}
\nabla^2\frac{1}{r^2}=-4\pi^2 \delta^4(\textbf{r}).
\end{equation}
\begin{eqnarray}
[-\nabla^2+4\pi^2
a^2\delta^4(\textbf{r})\frac{\partial^2}{2\partial
r^2}r^2]\varphi&=&4\pi^2a^2\delta^4(\textbf{r})\nonumber\\&&
\times[\frac{\partial^2}{2\partial r^2}r^2-1],
\end{eqnarray}
and
\begin{equation}
\varphi=1-\frac{a^2}{r^2}.
\end{equation}
The pseudopotential now becomes
\begin{equation}
V=8\pi^2
a^2\delta^4(\textbf{r})\frac{\partial^2}{2(\partial{r})^2}r^2
\nonumber  \hspace{4cm} \text{(PP4)}
\end{equation}
and
\begin{equation}
V'=4\pi^2a^2\int
d^4\textbf{r}\psi^*(\textbf{r})\psi^*(\textbf{r})\psi(\textbf{r})\psi(\textbf{r})
\end{equation}
We can now follow the steps in LHY to arrive at the energy $E_0$:
\begin{equation}
E_0=4\pi^2 a^2\rho N-\sum{}'[k^2+k_0^2-k
\sqrt{k^2+2k_0^2}-\frac{k_0^4}{2k^2}],
\end{equation}
where now
\begin{equation}
k_0=\sqrt{8\pi^2\rho a^2}.
\end{equation}
The summation in (9) converts into an integral:
\begin{equation}
N 64 \pi^4 a^6 \rho^2 \int_0^{\infty} \xi^2 d\xi[1+\xi^2-\xi
\sqrt{\xi^2+2}-\frac{1}{2\xi^2}],
\end{equation}
where $k=k_0\xi$.  This last integral has a \emph{ultra violet
divergence}:

\begin{equation}
\int\cong -\int^{\infty}\frac{1}{2}\frac{d\xi}{\xi}=-\frac{1}{2}ln
\xi|^{\infty}\cong-\frac{1}{2}ln k|^{\infty}.
\end{equation}
The cutoff for large k is $\sim\frac{1}{a}$. Thus the integral is
equal to $-\sim\frac{1}{2}ln\frac{1}{k_0 a}$.  We have thus
\begin{equation}
\frac{E_0}{N}=4\pi^2 a^2\rho +32 \pi^4 a^6 \rho^2|lnk_0a|+ \cdots.
\end{equation}
The depletion of the $\textbf{k} = 0$ state for dimension 3 was
given in LHY:
\begin{equation}
\langle n_{\textbf{k}=0}\rangle=N
[1-\frac{8}{3\sqrt{\pi}}\sqrt{\rho a^3}+\cdots].\nonumber
\hspace{2.5cm}[40b]
\end{equation}
One arrives at a similar expression for dimension 4:
\begin{equation}
\langle n_{\textbf{k}=0}\rangle=N [1-\beta \rho a^4 |ln \rho
a^4|+\cdots].
\end{equation}
where the logarithmic factor has the same origin as in (13) above.
Here $\beta$ is a numerical coefficient. Comparing equations (1)
and [40b] for dimension 3 with equations (13) and (14) for
dimension 4, we see that the expansion parameter changes from
dimensions 3 to 4 as follows:
\begin{equation}
\sqrt{\rho a^3}\rightarrow \rho a^4 |ln\rho a^4|.
\end{equation}

\section{DIMENSION 5}
Generalization of these perturbation calculations to dimension 5
proceeds seemingly without problem, but what replaces (14) becomes
\begin{equation}
\langle n_{\textbf{k}=0}\rangle=N [1-\infty]
\end{equation}
where $\infty$ represents a \emph{linear divergence} at large
\textbf{k}. This means that the single particle state
$\textbf{k}=0$ will be totally depleted by converting into many
pairs (\textbf{k}, -\textbf{k}) of single particle states. In
other words for interacting systems the BEC in dimensions 5  has
\cite{cnyang} basic group of 2 instead of 1, unlike the BEC of
dimensions 3 and 4. The details of this BEC need further study.

\section{DIMENSION 2}
In dimension 2 what replace (2) and (4) for dimension 3 are
\begin{equation}
\nabla^2(-lnr)=-2\pi \delta^2(\textbf{r}).
\end{equation}
and
\begin{equation}
\varphi=ln\frac{r}{a}.
\end{equation}
$\varphi$ is the correct scattering amplitude, but it is
impossible to generalize (3) to dimension 2 because unlike in
dimensions 3 and 4, (18) does not approach a limit as
$r\rightarrow \infty$.  However it \emph{changes very slowly} at
large \textbf{r}. To approximately solve this problem we search
for a $V$ which satisfies approximately
$-\nabla^2\varphi\cong-V\varphi$. Using (17) and (18) this
equation become
\begin{equation}
-2\pi \delta^2(\textbf{r})\cong -V[ln(r/a)].
\end{equation}
Now for a dilute system, the particles are at a large distance of
order
\begin{equation}
\textbf{R}=\sqrt{\rho}^{-1}.
\end{equation}
from each other.  Thus we may put $r\cong \textbf{R}$ in the RHS
of (19) and get
\begin{equation}
-2\pi \delta^2(\textbf{r})\cong -V[ln(R/a)].\nonumber
\end{equation}
or
\begin{equation}
V\cong\frac{1}{ln(R/a)}2\pi
\delta^2(\textbf{r})=\frac{4\pi\delta^2(\textbf{r})}{|ln(\rho
a^2)|}.\nonumber \hspace{2cm}\text{(PP2)}
\end{equation}
Here we take advantage of the fact that the logarithm varies very
slowly with its argument if the latter is large.  (The above
derivation is not rigorous but I believe it to be correct.)

With (PP2) we proceed as in LHY and obtain
\begin{equation}
E_0=\frac{4\pi \rho N}{|ln \rho a^2|}-\sum{}'[k^2+k_0^2-k
\sqrt{k^2+2k_0^2}]
\end{equation}
where \begin{equation} k_0^2=8\pi \rho/|ln\rho a^2|.
\end{equation}

\begin{table*}
\begin{tabular} {|l|c|c|l|} \hline
Dimension  & Pseudopotential for fixed scatterer &$E_0/N$&
Comments\\\hline 1& --- & $\pi^2\rho^2/3(1-\rho a)^2$ & No BEC
\\ \hline
2 &  {\large$\frac{\sim 4\pi\delta^2}{|ln(\rho a^2)|}$ }  & \large{$\sim \frac{4\pi \rho}{|ln(\rho a^2)|}+\cdots$} & Almost BEC\\
 \hline
3 & {\normalsize $8\pi a \delta^3 \frac{\partial}{\partial r}r$  }
& {\normalsize $4\pi \rho a(1+\frac{128\sqrt{\rho
a^3}}{15\sqrt{\pi}}+\cdots)$} &BEC. BG = 1 Boson
\\ \hline
4& \normalsize{$8\pi^2 a^2 \delta^4\frac{\partial^2}{2(\partial
r)^2}r^2$} & {$ 4\pi^2 \rho a^2(1+4\pi^2 \rho a^4|ln(\rho
a^4)|+\cdots) $} &BEC. BG = 1 Boson
\\ \hline  5  & &  ? &BEC.   BG = pair of Bosons \\
\hline
\end{tabular}
\caption{\label{tab:table1}Comparison of BEC in dilute interacting
Boson gases for different dimensions. (BG = basic group [6])}
\end{table*}
There is no subtraction term in (21), unlike [23], because now
$V'=V$. The summation gives a contribution to $E_0/N$:
\begin{equation}
-\frac{16\pi\rho}{A^2}lnA \nonumber
\end{equation}
where $A=|ln\rho a^2|$. But it is not clear that this term is
reliable.

We summarize the results for dimension $n$ in Table I for dilute
interacting Boson systems.

With incredible new technology it has been possible to study
experimentally 2D and 1D systems \cite{AG}. Perhaps the
qualitative and quantitative features exhibited in Table I above
for different dimensions could be studied with these new
technologies and with powerful computers.

\end{document}